\documentclass[aps,pra,twocolumn]{revtex4}
\usepackage{braket}
\usepackage{bm}
\usepackage{amsmath}
\usepackage{enumitem}
\usepackage{graphicx}
\usepackage{graphics}
\usepackage[
	pdftex,
	colorlinks=true,
  	urlcolor=blue,
  	linkcolor=blue,
  	citecolor=blue
]{hyperref}
\usepackage{xcolor}
\begin{document}


\title{Renormalization-group study of the many-body localization transition in one dimension}



\author{Alan Morningstar}
\affiliation{Department of Physics, Princeton University, Princeton, New Jersey 08544, USA}

\author{David A. Huse}
\affiliation{Department of Physics, Princeton University, Princeton, New Jersey 08544, USA}


\date{\today}

\begin{abstract}
Using a new approximate strong-randomness renormalization group (RG), we study the many-body localized (MBL) phase and phase transition in one-dimensional quantum systems with short-range interactions and quenched disorder.  Our RG is built on those of Zhang {\it et al.}~\cite{Zhang-Huse2016} and Goremykina {\it et al.}~\cite{Goremykina-Serbyn2018}, which are based on thermal and insulating blocks. Our main addition is to characterize each insulating block with two lengths: a physical length, and an internal decay length $\zeta$ for its effective interactions.  In this approach, the MBL phase is governed by a RG fixed line that is parametrized by a global decay length $\tilde{\zeta}$, and the rare large thermal inclusions within the MBL phase have a fractal geometry. As the phase transition is approached from within the MBL phase, $\tilde{\zeta}$ approaches the finite critical value corresponding to the avalanche instability, and the fractal dimension of large thermal inclusions approaches zero. Our analysis is consistent with a Kosterlitz-Thouless-like RG flow, with no intermediate critical MBL phase.
\end{abstract}


\maketitle

\section{Introduction\label{sec:Introduction}}

Most interacting many-body quantum systems, when isolated from any external environment and evolving under their own unitary quantum dynamics, act as a heat ``bath'' for themselves, allowing their subsystems to reach thermal equilibrium. The main generic exception to this is systems that are many-body localized.  Such MBL systems can locally preserve locally encoded information indefinitely, and their long-time states can fail to conform to equilibrium statistical mechanics~\cite{Anderson1957,Basko-Altshuler2006,Oganesyan-Huse2007,Nandkishore-Huse2014a}. The existence of a many-body localized phase in certain one-dimensional systems with strong quenched randomness and weak short-range interactions has been proven under certain minimal assumptions~\cite{Imbrie2016}.  However, this MBL phase is stable only for strong enough disorder, and as the disorder strength is reduced there is a nonequilibrium dynamical phase transition to the thermal phase where the system does constitute a ``bath'' that can bring itself to thermal equilibrium.  The thermal phase is believed to obey the eigenstate thermalization hypothesis (ETH), where all eigenstates of the dynamics are at thermal equilibrium and are thus volume-law entangled, and late-time properties are governed by equilibrium statistical mechanics for any initial state~\cite{Deutsch1991,Srednicki1994,Rigol-Olshanii2009,Dalessio-Rigol2016}. Conversely, the MBL phase at strong disorder has area-law-entangled eigenstates of the dynamics, logarithmically spreading quantum correlations, and emergent local integrals of motion (``l-bits'')~\cite{Znidaric-Prelovsek2008,Pal-Huse2010,Bardarson-Moore2012,Serbyn-Abanin2013a,Serbyn-Abanin2013b,Huse-Oganesyan2014,Serbyn-Abanin2015}.

The nonequilibrium dynamical phase transition between the MBL and thermal phases, henceforth referred to as the MBL transition, is unconventional in that it is an eigenstate phase transition occurring at non-zero energy densities; it can even occur at infinite temperature~\cite{Nandkishore-Huse2014a,Abanin-Serbyn2018}.  This transition has been studied via exact diagonalization (ED), however those studies all show significant finite-size effects that prevent reliable extrapolation to the asymptotic critical behavior~\cite{Pal-Huse2010,Kjall-Pollmann2014,Luitz-Alet2015,Luitz-Alet2016,Serbyn-Moore2016,Yu-Clark2016,Luitz2016,Luitz-Lev2017,Khemani-Huse2017a,Khemani-Huse2017b,Geraedts-Nandkishore2017,Gray-Bayat2018,Herviou-Bardarson2018,Roy-Luitz2018}. In fact, all but one of such studies indicate an apparent correlation length critical exponent which violates the bound $\nu \geq 2$ for disordered, one-dimensional systems~\cite{Chandran-Oganesyan2015}. 

In order to investigate the asymptotic critical behavior of the MBL transition, strong-randomness renormalization group (RG) approximations have been devised and applied~\cite{Vosk-Altman2015,Potter-Prameswaran2015,Dumitrescu-Potter2017,Thiery-DeRoeck2017a,Thiery-DeRoeck2017b,Zhang-Huse2016,Goremykina-Serbyn2018,Dumitrescu-Vasseur2018}.  The earliest of these works~\cite{Vosk-Altman2015,Potter-Prameswaran2015,Dumitrescu-Potter2017} provided numerically implemented RGs designed to capture the physics of interactions between locally thermal and MBL regions in systems containing thousands or more of such subsystems.  Each assuming a one-parameter scaling ansatz, apparent exponents near $\nu \cong 3$ were found.  Following the exposition of the \textit{avalanche} mechanism as a possible scenario for the MBL transition~\cite{DeRoeck-Huveneers2017,Luitz-DeRoeck2017}, a microscopically motivated RG based on ETH and perturbative diagonalization, and containing the physics of quantum avalanches, was developed in~\cite{Thiery-DeRoeck2017a,Thiery-DeRoeck2017b}. This also yielded $\nu\cong 3$ under a power-law scaling ansatz, and numerical evidence suggesting that the critical point is, in some senses, fully many-body localized.  Although these papers did not explicitly suggest a Kosterlitz-Thouless (KT) type RG flow, the ingredients for such a flow are present in their RG in a way that is essentially the same as in our RG~\cite{Thiery-DeRoeck2017a,Thiery-DeRoeck2017b}.

Along a parallel line of developments, an oversimplified but analytically tractable RG designed to try to capture some aspects of the transition was solved in~\cite{Zhang-Huse2016}.  This ``toy'' RG has a critical fixed point that does obey one-parameter scaling, but it is physically incorrect in having a spurious symmetry between the MBL and thermal phases.  A modification of this toy RG that does not have this incorrect symmetry, but is still somewhat analytically tractable was developed and investigated in~\cite{Goremykina-Serbyn2018}, and a two-parameter KT-like RG flow was found.  Subsequent work~\cite{Dumitrescu-Vasseur2018} argued that a KT-like RG flow follows generally from considering an MBL transition driven by avalanches. This work also revisited the older RGs of~\cite{Vosk-Altman2015,Dumitrescu-Potter2017}, showing that they are consistent with a KT-like two-parameter RG flow.  This is likely true of most of the RGs that were fit to one-parameter scaling forms yielding exponents near $\nu\cong 3$, and we show below that our new RG also gives a similar estimate of $\nu$ when fit to one-parameter scaling.  

Two of these recent papers have also suggested that there may be an intermediate ``critical" phase that is MBL, where the lengths of locally thermal inclusions within this intermediate MBL phase are distributed according to a power law, and the exponent of this power law varies along the KT-like RG fixed line~\cite{Goremykina-Serbyn2018,Dumitrescu-Vasseur2018}.  As we report below, we instead find that the distribution of the lengths of locally thermal regions is a power law in the limit of large lengths only at the critical point, so there is no intermediate phase.  Our fixed line is parametrized by the exponential decay length $\tilde{\zeta}$ for the effective spin-flip interactions, as proposed in Ref.~\cite{Dumitrescu-Vasseur2018}.

In this paper we build on the work of~\cite{Zhang-Huse2016} and~\cite{Goremykina-Serbyn2018} to define a RG description of the MBL transition and the MBL phase which is more detailed, and that captures more of the correct physics.  Due to extra features in our RG, it is not fully analytically solvable, however it is somewhat tractable with a combination of analytic and numerical methods. In Section~\ref{sec:TheRG} we define our RG, detailing coarse-grained variables and their behavior under RG transformations. The structure of the RG flow, the character of thermal inclusions, and the question of whether or not an intermediate critical MBL phase is supported by our coarse-grained description are addressed in Sections~\ref{sec:Numerics} and~\ref{sec:Analytics} via numerical and analytic methods. In Section~\ref{sec:Conclusions} we collect and organize the implications of our model.	

\section{The RG \label{sec:TheRG}}
We consider a coarse-grained description of disordered Hamiltonian or Floquet spin chains similar to the simplified RGs analyzed in~\cite{Zhang-Huse2016,Goremykina-Serbyn2018}, where locally thermalizing and MBL segments of the chain are assumed to be sharply distinct from one another and are characterized by just one and two, respectively, coarse-grained properties. Such segments are referred to as $T$ and $I$ \textit{blocks}, respectively.  Two adjacent blocks of the same type can always be combined to form a larger block of that type, so the blocks within a chain alternate between being $T$ and $I$. An $I$ block is a contiguous segment of the system that, when isolated from the rest of the chain, has a complete set of localized integrals of motion (l-bits) that are localized on a length scale smaller than the length of the $I$ block.  A $T$ block similarly represents a part of the system that, when isolated, has eigenstates of the dynamics that have near-thermal entanglement; there are no l-bits within a $T$ block that are localized on any length scale shorter than the $T$ block itself. Interactions between the blocks can be viewed as being between the l-bit degrees of freedom within an $I$ block and the non-local degrees of freedom of a $T$ block.  When an $I$ block is coupled to an adjacent $T$ block, the resulting interactions flip l-bits at a rate that falls off exponentially with the distance between the l-bit and the edge of the $T$ block.

This type of RG is treating the limit in which systems are so large that $T$ blocks of all sizes are present, however spatially rare they may be.  For smaller systems this may not be the case, and the apparent MBL transition in such small finite-size systems may be more attributable to the first appearance of such entangled $T$ regions, rather than the physics captured by our RG~\cite{Gopalakrishnan-Huse2019}.

The crucial strategy of breaking the spurious $I$-$T$ symmetry of the RG of Ref.~\cite{Zhang-Huse2016} in order to capture more of the correct physics of the MBL transition was introduced by Goremykina {\it et al.} \cite{Goremykina-Serbyn2018}. Following this strategy, the key new ingredient that we add to the RGs of Refs.~\cite{Zhang-Huse2016,Goremykina-Serbyn2018} is that, in addition to $I$ and $T$ blocks having physical lengths $L^I$ and $L^T$, respectively, each $I$ block also is assumed to have an internal decay length $\zeta$ which governs the exponential decay of the l-bit-flip interactions that were discussed earlier in this section.  We normalize this decay length so that $\zeta=1$ is the threshold for the avalanche instability~\cite{DeRoeck-Huveneers2017,Luitz-DeRoeck2017}; therefore all $I$ blocks have $\zeta<1$. We further define two more lengths for $I$ blocks which are useful in formulating the RG: If $I$ block $n$ has physical length $L_n^I$ and decay length $\zeta_n$, we thus define its effective interaction length as $l_n=L_n^I/\zeta_n$, which is related to the logarithm of the interactions across the $I$ block. We also define the length $d_n=l_n-L^I_n$, so $d_n$ is a length quantifying how close $I$ block $n$ is to the avalanche instability. All $I$ blocks have $L^I < l$ and $d>0$, and the avalanche instability occurs when $l=L^I$ and $d=0$ (both a result of $\zeta=1$). Any two of the four lengths $L^I$, $\zeta$, $l$, and $d$ can be used to fully characterize an $I$ block, and only $L^T$ is needed to specify a $T$ block. If $I$ block $n$ interacts with a single $T$ block of length $L^T_{n-1}$, and that single $T$ block is only in contact with this $I$ block, then the $T$ block starts an avalanche across the $I$ block, and this  avalanche will halt within the $I$ block only if $L^T_{n-1} < d_n$.  If $L^T_{n-1} \geq d_n$ then the avalanche traverses the entire $I$ block, so the $T$ block fully thermalizes the $I$ block.  Thus $d_n$ is the minimum length of a $T$ block that can fully thermalize $I$ block $n$ by itself.

A key feature of the two previous approximate RGs of this type~\cite{Zhang-Huse2016,Goremykina-Serbyn2018} is that the fixed point distributions are product distributions over the blocks: no interblock correlations get generated by the RG.  This simplicity allows useful analytic results about the RG to be obtained.  This is a feature we also preserve in formulating our RG.  The basic RG steps in all three of these RGs are block decimation \textit{moves} that either replace a 3-block sequence $ITI$ with one longer $I$ block, or replace a sequence $TIT$ with one longer $T$ block.  To prevent the production of any correlations in the joint probability distribution over blocks, the decimations that occur are determined only by properties of the central block of the 3-block sequence that is decimated. The detailed RG rules for carrying out these 3-block moves is provided in the following section.

\subsection{RG rules \label{subsec:RGrules}}
A sliding RG scale, or \textit{cutoff}, is denoted by $\Lambda$.  This cutoff length steadily grows as the RG runs. A $T$ block is decimated by a 3-block RG move when the cutoff length reaches its physical length $L^T$, as in \cite{Zhang-Huse2016}, and an $I$ block is decimated when the cutoff length reaches length $d$. Since these lengths play a similar role in our RG, we will call them the ``primary length'' of the corresponding block: For a $T$ block the primary length is the physical length $L^T$, while for an $I$ block the primary length is $d$, which is the length of the shortest $T$ block that can, by itself, thermalize that $I$ block. We assume that the primary lengths are continuously distributed, so the order of RG moves is unambiguous.

\begin{figure}
	\includegraphics[width=0.9\linewidth]{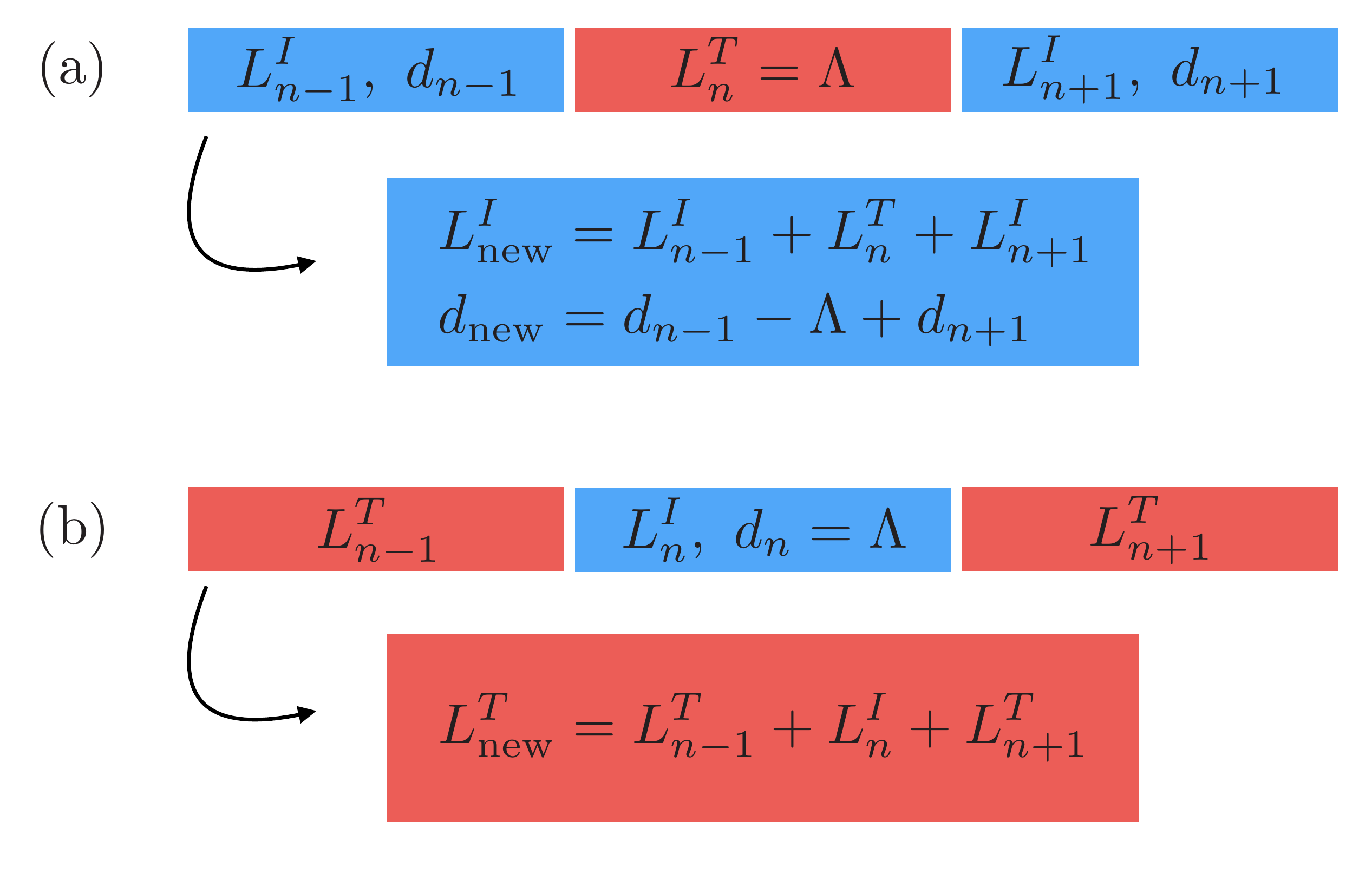}
	\caption{ A summary of the RG rules. The rules are formulated in terms of the lengths $L^I$ and $d$ for $I$ blocks and $L^T$ for $T$ blocks. (a) An $ITI\mapsto I$ move where a $T$ block at the cutoff ($L^T_n=\Lambda$) fails to thermalize the neighboring $I$ blocks. (b) A $TIT\mapsto T$ move where an $I$ block at the cutoff ($d_n=\Lambda$) thermalizes via interactions with neighboring $T$ blocks. \label{fig:rules}}
\end{figure}

An $ITI\mapsto I$ RG move represents the physical scenario where the central thermal block $n$ is too small a bath to thermalize the adjacent blocks, which are $I$ blocks $n-1$ and $n+1$, so the local quantum avalanches started in those $I$ blocks by this $T$ block both halt~\citep{Thiery-DeRoeck2017b,Luitz-DeRoeck2017}.  The $T$ block is thus localized, and we combine all three blocks to make one new larger $I$ block whose physical length is simply the sum:
\begin{equation}
L^I_\mathrm{new} = L^I_{n-1} + \Lambda + L^I_{n+1}~.
\end{equation} 
Our RG does this move when the cutoff $\Lambda$ equals the length $L_n^T$ of the $T$ block, as is noted in the above equation.  The minimum total length of $T$ block that would be needed to thermalize both of these $I$ blocks is $d_{n-1}+d_{n+1}$.  The $T$ block has provided some of that, but not enough, so the primary length of the new $I$ block is 
\begin{equation}
	\label{eqn:dInew}
	d_\mathrm{new} = d_{n-1} - \Lambda + d_{n+1}~,
\end{equation}
which is the remaining $T$ block length that would be needed to thermalize this new $I$ block.  Note that since at this point in the RG $d_i>\Lambda$ for any $I$ block $i$, this remains true for the new $I$ block: $d_\mathrm{new}>\Lambda$.  The other lengths for this new $I$ block are then given by $l_{\mathrm{new}}=L^I_\mathrm{new} + d_\mathrm{new}=l^I_{n-1}+l^I_{n+1}$ (note that this update rule for $l$ is the same as that used in Ref.~\cite{Goremykina-Serbyn2018}), and $\zeta_{\mathrm{new}}=L^I_\mathrm{new}/l_\mathrm{new}<1$. The rules for $ITI\mapsto I$ moves are summarized in Figure~\ref{fig:rules}\hyperref[fig:rules]{(a)}.

A $TIT\mapsto T$ move represents the complementary scenario where the $I$ block is too weakly insulating to prevent the two adjacent $T$ blocks from being entangled with each other in the eigenstates of the system's dynamics---the two avalanches started by the two $T$ blocks meet and merge in the middle of the $I$ block.  The $I$ block is thus thermalized, and we combine all three blocks to make one new larger $T$ block whose physical length is simply the sum:
\begin{equation}
L^T_\mathrm{new} = L^T_{n-1} + L^I_{n} + L^T_{n+1}~.
\end{equation} 
Our RG does this move when the cutoff $\Lambda$ equals the primary length $d_n$ of the central $I$ block. An important feature of our RG is the possibility for $L^I_{n}$ to be large even though $d_n = \Lambda$. This corresponds to the scenario where a large insulating region that is near the avalanche instability $\zeta=1$ is thermalized. The rules for $TIT\mapsto T$ moves are summarized in Figure~\ref{fig:rules}\hyperref[fig:rules]{(b)}.

These RG rules are chosen as a minimal modification of those of Refs.~\cite{Zhang-Huse2016,Goremykina-Serbyn2018} that puts in the physics of the avalanche instability of the MBL phase.  They are chosen to preserve the property that the RG does not generate interblock correlations, which allows us to obtain some analytic results about this RG.  The RG moves that happen are those associated with the avalanches due to $T$ blocks with length equal to the cutoff $\Lambda$.  If those avalanches are halted by both of the adjacent $I$ blocks, that results in an $ITI\mapsto I$ move.  If an $I$ block can be thermalized by a $T$ block of length $\Lambda$, then the two adjacent $T$ blocks do indeed thermalize this $I$ block, and that results in a $TIT\mapsto T$ move.

We do not claim that this still-``simple'' RG does include all of the physics of this Kosterlitz-Thouless-like universality class of MBL transition.  But we have put in more of the physics of the MBL phase, and as a result we obtain a RG fixed line for that phase that is of a different character from that described in Ref.~\cite{Goremykina-Serbyn2018}.  

One piece of the correct physics that we still leave out is the internal dynamics within the $T$ blocks.  Because of this, our RG as we have described it so far does not contain the physics of rare insulating-like ``bottlenecks'' within the thermal phase~\cite{Agarwal-Demler2015,Gopalakrishnan-Knap2016,Agarwal-Knap2017}.  This can be remedied by including an entanglement time for each new $T$ block.  This time is set by the $I$ bottleneck that is decimated in the $TIT\mapsto T$ move that produces the new $T$ block.  
We are assuming that the time scales of the physics of any RG moves that occur later in the RG are much longer than this time, so it is assumed not to feed back and alter the above RG rules and we can thus ignore this additional property of the $T$ blocks.  More elaborate RGs of this type that include this physics more accurately can be investigated in the future.

\subsection{Single-block probability distributions \label{subsec:Singleblock}}
An infinite system can be specified by single-block probability distributions $\rho^T_\Lambda(L^T)$ and $\rho^I_\Lambda (d,L^I)$ that flow with the cutoff $\Lambda$~\citep{Zhang-Huse2016,Goremykina-Serbyn2018}. This is valid because the RG rules do not introduce interblock correlations.  Note that in our RG each $I$ block has a nontrivial joint probability distribution for its two lengths. It is useful to also define the marginal distribution, $\mu^I_\Lambda (d) \equiv \int_0^\infty \rho^I_\Lambda (d,L^I) dL^I$, of $I$ blocks because $\rho^T_\Lambda (\Lambda)$ and $\mu^I_\Lambda (\Lambda)$---the density of $T$ and $I$ blocks with their primary length at the cutoff---determine the rate of 3-block moves as the RG proceeds.  For notational brevity, the subscript $\Lambda$ indicating cutoff-dependence is omitted unless required for clarity.

\section{Numerics \label{sec:Numerics}}
The simplest method to investigate this RG with is direct numerical simulation of finite systems. Once a chain of $N$ blocks is initialized according to a bare model, the RG proceeds by repeating the steps
\begin{enumerate}
	\label{enum:RG}
	\item find the block with the smallest primary length,
	\item decimate this block in an $ITI\mapsto I$ or $TIT\mapsto T$ move.
\end{enumerate}
This process stops when there is only one of each type of block left. At that point, the block with the larger primary length determines the phase of the finite-size system. 

The data shown in this paper was generated by running the RG on finite systems initialized in the following way:  A chain of blocks has alternating $T$ and $I$ blocks.  The lengths of bare $T$ blocks were sampled from the distribution $\rho^T(L^T)=\exp(1-L^T)$, with $L^T \in [1,\infty)$. We chose to give all $I$ blocks the same initial $\zeta$, and we use this $\zeta_\mathrm{bare}$ as the tuning parameter of the transition. In order to ensure that in the bare model there were both types of blocks near the cutoff, we sampled the length $d$ of $I$ blocks, with $d \in [1,\infty)$, from $\mu^I(d)=\lambda(\zeta_\mathrm{bare})\exp ( (1-d)\lambda(\zeta_\mathrm{bare}))$, with $\lambda(\zeta_\mathrm{bare})$ a function of $\zeta_\mathrm{bare}$ chosen so that $\braket{L^I} = \braket{L^T}$ up to statistical fluctuations.  This bare model was chosen for simplicity, and to allow a significant fraction of $TIT\mapsto T$ moves to occur in order to quickly wash away any bias due to the bare model. The results at large RG scales are not qualitatively different upon varying the family of bare models.

\subsection{One-parameter scaling \label{subsec:Oneparameter}}
As mentioned earlier, before a KT-like scenario for the MBL transition was proposed~\cite{Goremykina-Serbyn2018,Dumitrescu-Vasseur2018}, studies using approximate RG treatments of the MBL transition assumed a one-parameter scaling ansatz~\cite{Vosk-Altman2015,Potter-Prameswaran2015,Dumitrescu-Potter2017,Thiery-DeRoeck2017a,Thiery-DeRoeck2017b,Zhang-Huse2016}. These works reported apparent correlation-length critical exponents in the range $\nu \cong 2.5$ to $\nu \cong 3.5$. 

\begin{figure}
	\begin{minipage}{0.5\linewidth}
		\includegraphics[width=1.0\linewidth]{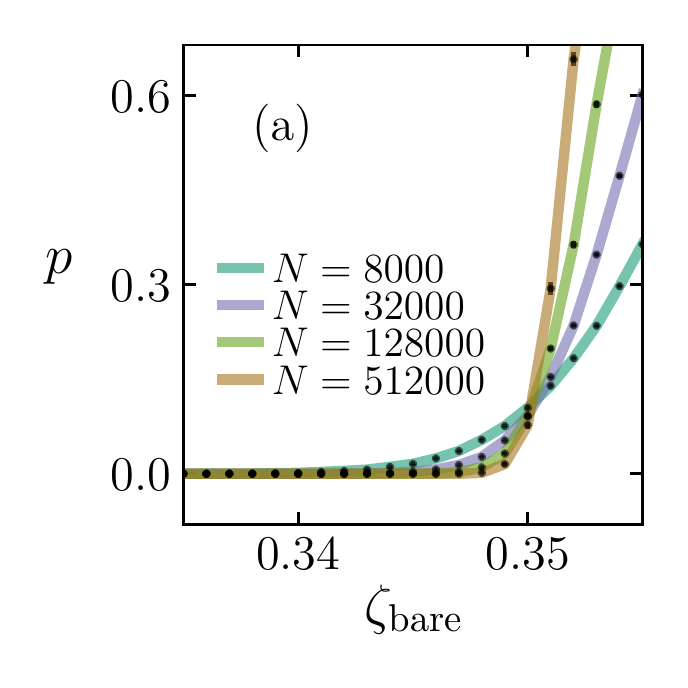}
	\end{minipage}\hfill
	\begin{minipage}{0.5\linewidth}
		\includegraphics[width=1.0\linewidth]{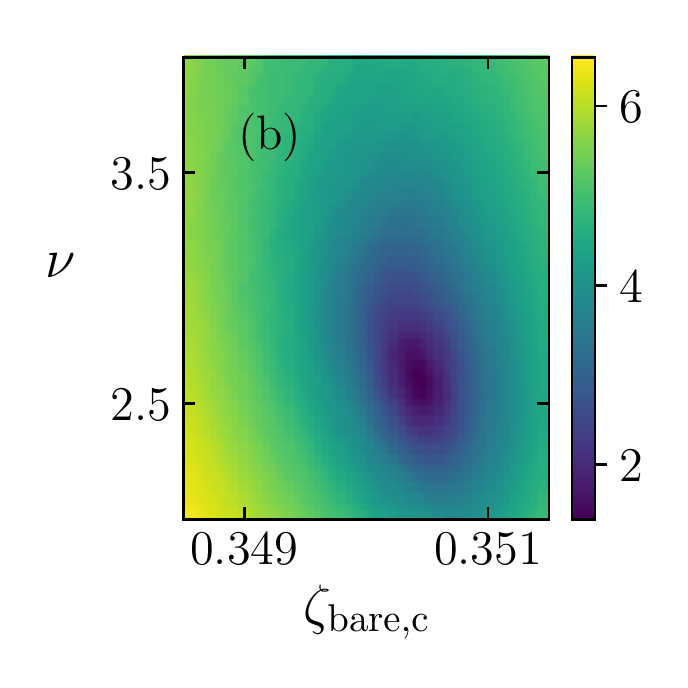}
	\end{minipage}
	\caption{(a) The probability of a sample being thermal. The curves are for $N$ from $8\times 10^3$ to $512\times 10^3$ by powers of four.  The number of samples used to generate each data point is $10^9/N$. Standard errors are smaller than the markers. (b) $\log Q$, an indicator of the quality of the one-parameter scaling ansatz for different values of $\nu$ and $\zeta_\mathrm{bare,c}$. Smaller values indicate a better fit, with $\log Q = 0$ indicating the best possible result. \label{fig:scaling}}
\end{figure}

In order to demonstrate consistency with these one-parameter-scaling results, before moving on to the two-parameter RG flow which we believe is actually present, we first assume that the probability of a sample being in the thermal phase is a function of the scaled bare decay length $(\zeta_\mathrm{bare} - \zeta_\mathrm{bare,c})N^{1/\nu}$, where $\zeta_{\mathrm{bare,c}}$ is the critical value~\cite{Thiery-DeRoeck2017a}. In Figure~\ref{fig:scaling}\hyperref[fig:scaling]{(a)} we plot this probability for different system sizes. Using the pyfssa package~\cite{Sorge2015}, the relative quality of such a scaling ansatz is assessed for different values of $\nu$ and $\zeta_\mathrm{bare,c}$~\cite{Gray-Bayat2018,Houdayer-Hartmann2004}. The result, shown in Figure~\ref{fig:scaling}\hyperref[fig:scaling]{(b)}, indicates $\zeta_\mathrm{bare,c} \cong 0.350$ and $\nu \cong 2.6$.  This is consistent with the bound $\nu\ge 2$ and the one-parameter scaling results of previous RG studies. 

We note that, upon closer inspection, the curves in Figure~\ref{fig:scaling}\hyperref[fig:scaling]{(a)} cross at smaller values of $p$ and $\zeta_\mathrm{bare}$ as $N$ increases; this was also observed in Ref.~\cite{Thiery-DeRoeck2017a}. Since Figure~\ref{fig:scaling}\hyperref[fig:scaling]{(b)} indicates smaller values of $\zeta_\mathrm{bare,c}$ fit with much larger values of $\nu$, a flow to larger $\nu$ may occur but is difficult to observe directly. If the RG flow of our model is truly a KT-like two-parameter flow, as we argue below, then $\nu$ is not truly finite and the value $\nu \cong 3$ is due to the severe finite-size effects generally present in systems exhibiting KT-like scaling.

\subsection{Two-parameter flow \label{subsec:Twoparameter}}
The idea that the MBL transition may not be accounted for by a one-parameter scaling theory was proposed in Ref.~\cite{Thiery-DeRoeck2017a} and elaborated on in Refs.~\cite{Goremykina-Serbyn2018,Dumitrescu-Vasseur2018}.  In particular, these latter works suggested the flow of a correct RG description is KT-like, with the critical point at the end of a MBL fixed line. 

In order to map the RG flow of our model, we introduce the fraction $f$ of the system's length that is in the $T$ blocks, and a running estimate of a global decay length $\tilde{\zeta}$ for l-bit-flip interactions across the entire system. These quantities are recorded as a function of the cutoff $\Lambda$ as the RG runs. The fraction of a system that is locally thermal is given by
\begin{equation}
	\label{eqn:f}
	f = \frac{\braket{L^T}}{\braket{L^T}+\braket{L^I}},
\end{equation}
where $\langle \cdot \rangle$ denotes an average over qualifying blocks at a given ``instant'' in the RG flow.  The global decay length $\tilde{\zeta}$ is calculated by considering the combination of all blocks according to the $ITI\mapsto I$ rules, and using the definition $\zeta = L^I / l$. This results in 
\begin{equation}
	\label{eqn:zetatilde}
	\tilde{\zeta} = \frac{\braket{L^I} + \braket{L^T}}{\braket{l}},
\end{equation}
which shows that $\tilde{\zeta}$ can take values $>1$, indicating a system which is beyond the avalanche instability of the MBL phase and thus is in the thermal phase.  By mapping the RG flow using $f$ and $\tilde{\zeta}$ we investigate the phases and phase transition of our model.

\begin{figure*}[t]
	\begin{minipage}{0.31\linewidth}
		\includegraphics[width=1.0\linewidth]{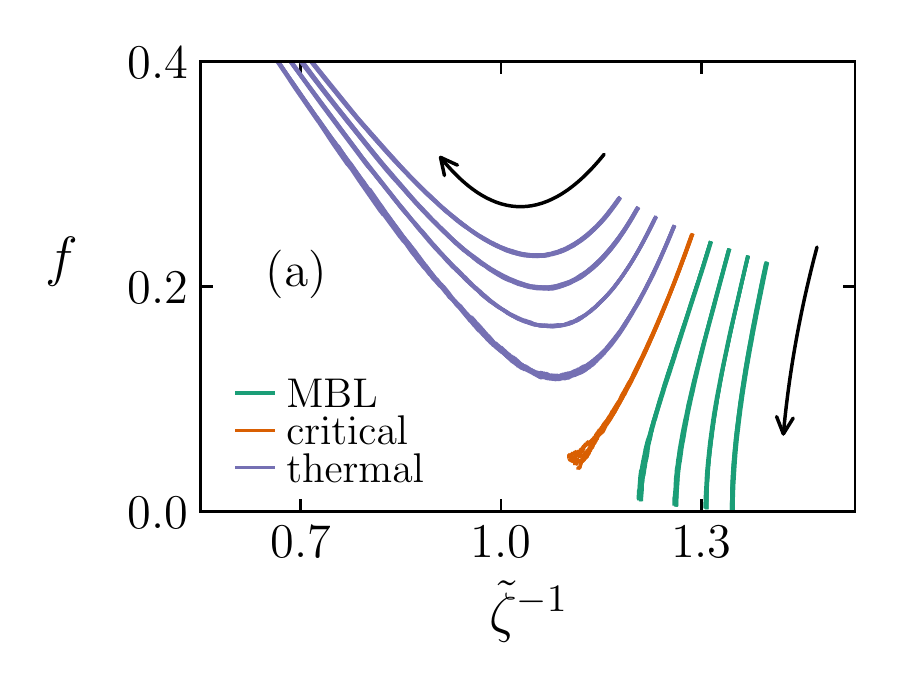}
	\end{minipage}\hfill
	\begin{minipage}{0.31\linewidth}
		\includegraphics[width=1.0\linewidth]{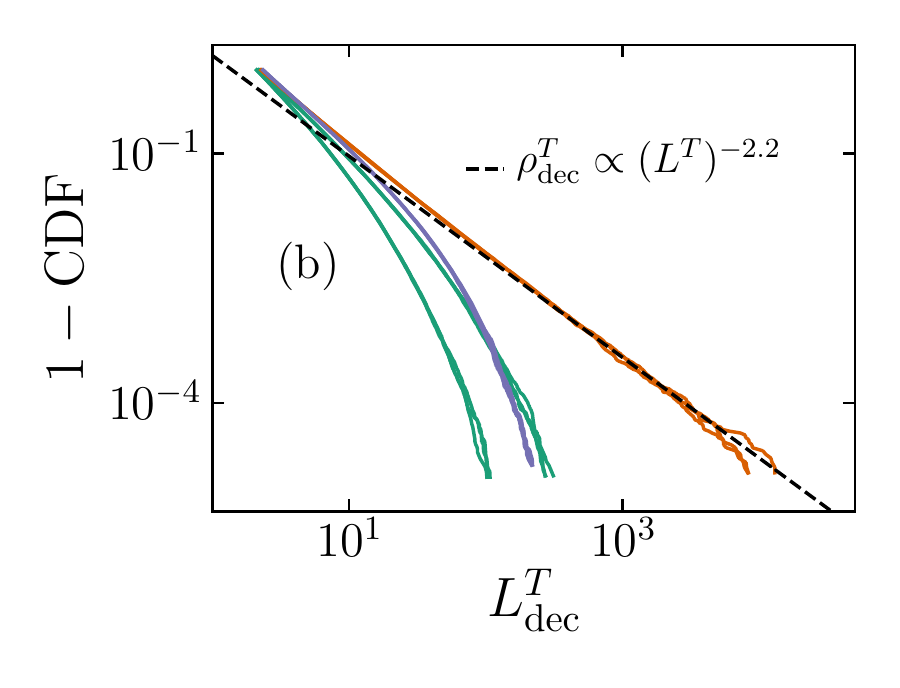}
	\end{minipage}
	\begin{minipage}{0.35\linewidth}
		\includegraphics[width=1.0\linewidth]{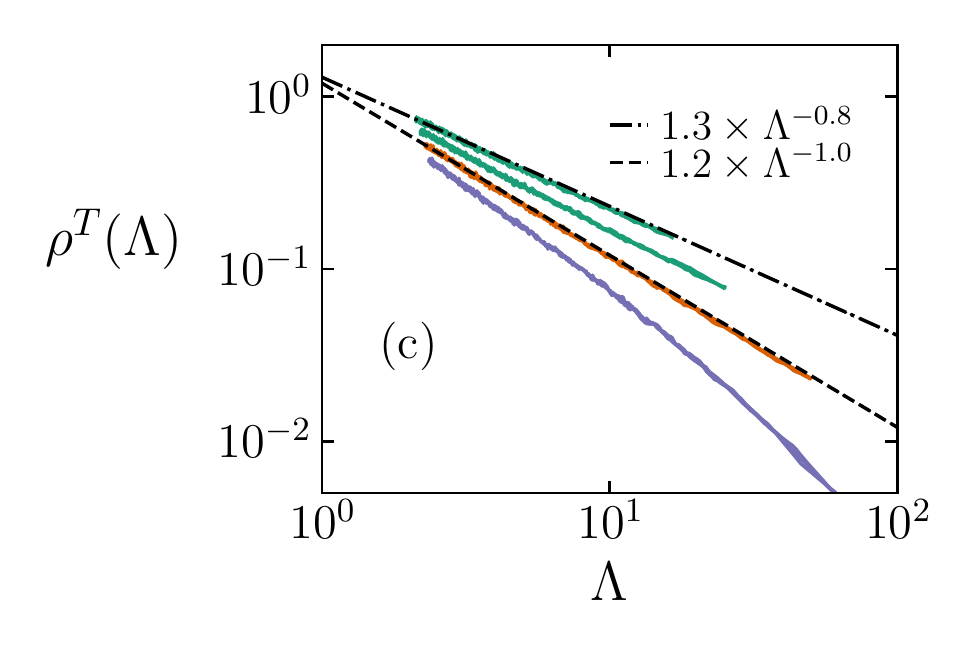}
	\end{minipage}\hfill
	\caption{Results of numerical simulations of single samples with $N=10^7$. Each thick line consists of three (b and c) to five (a) thin lines representing single samples: when these individual samples can be distinguished, that gives a crude estimate of the statistical errors. The data shown was recorded during the final 10\% of RG steps so as to isolate the large-$\Lambda$ behavior. (a) The RG flow plotted on the same plane as that of~\cite{Dumitrescu-Vasseur2018} for direct comparison. Simulations were done with $\zeta_\mathrm{bare} \in [0.330,0.370]$ (right to left) by steps of 0.005. (b) The distribution of the physical lengths of decimated thermal blocks at criticality and in each phase for $\zeta_\mathrm{bare} = [0.300,0.375]$ by steps of 0.025. (c) The scaling of $\rho^T(\Lambda)$ with $\Lambda$ for $\zeta_\mathrm{bare} \in [0.300,0.375]$ again (top to bottom). $\rho^T(\Lambda)$ was calculated by numerically estimating the derivative of the cumulative distribution function of $L^T$ at $L^T=\Lambda$ with a quadratic fit to the first $10^3$ data points. This was done up to the scale where $10^4$ $T$ blocks remained. The raw data was smoothed using a Savitzky-Golay filter with a window containing 1\% of the data. \label{fig:flow}}
\end{figure*}

Similar to the phenomenological RG flow shown in~\cite{Dumitrescu-Vasseur2018}, we map the RG flow of our model in the $(\tilde{\zeta}^{-1},f)$-plane by simulating the RG on finite systems and monitoring the flow of $f$ and $\tilde{\zeta}$ with $\Lambda$. The result, shown in Figure~\ref{fig:flow}\hyperref[fig:flow]{(a)}, indicates that the MBL phase is a fixed line at $f=0$ which can be parametrized by $\tilde{\zeta}_\infty \in (0,1)$, the global decay length as $\Lambda \to \infty$. This flow is consistent with the KT-type scenario~\cite{Goremykina-Serbyn2018,Dumitrescu-Vasseur2018}, where the critical fixed point is a terminus of the MBL fixed line located at $f=0$, $\tilde{\zeta}=1$, and the transition is driven by the avalanche instability. However, finite-size effects limit the extent to which numerical simulations alone can confirm this.

\subsection{Distribution of thermal inclusions \label{subsec:Thermalinclusions1}}

Many approximate RG studies~\cite{Thiery-DeRoeck2017a,Dumitrescu-Potter2017,Dumitrescu-Vasseur2018,Goremykina-Serbyn2018} and one recent ED study~\cite{Herviou-Bardarson2018} of the MBL transition seem to agree on the prediction that physical lengths of thermal inclusions in critical systems are distributed according to a power law  $\propto (L^T)^{-\alpha}$, with the exponent taking an apparently universal value near $\alpha = 2$. In addition, and consistent with a KT-type scenario, is the possibility of an intermediate critical MBL phase~\cite{Dumitrescu-Vasseur2018} where the lengths of thermal inclusions are power-law distributed with a continuously varying exponent \cite{Goremykina-Serbyn2018,Dumitrescu-Vasseur2018,Herviou-Bardarson2018}. 

We investigate the distribution of the physical lengths of thermal inclusions in MBL and critical systems by interpreting $T$ blocks that are decimated in $ITI\mapsto I$ moves during the RG to be representative of the thermal inclusions in large MBL regions. The length distribution of these decimated $T$ blocks, herein denoted as $\rho^T_\mathrm{dec}(L^T_\mathrm{dec})$, has a tail to large lengths that is at least as heavy as that of the \textit{instantaneous} distribution $\rho^T (L^T)$. The tails of the two distributions take the same form in the case where after the RG scale reaches some value, only $ITI\mapsto I$ moves occur.  In Figure~\ref{fig:flow}\hyperref[fig:flow]{(b)} we plot the complimentary cumulative distribution function ($1-\mathrm{CDF}$) of this distribution $\rho^T_\mathrm{dec}(L^T_\mathrm{dec})$ for systems at criticality and in the MBL and thermal phases. This shows that our RG yields power-law-distributed large thermal inclusions only near criticality, with an exponent $\alpha \cong 2.2$ at large $L^T_\mathrm{dec}$.  It is possible that this estimate of $\alpha$ would change if even larger systems were accessible, since our RG still exhibits some finite-size effects.  Of course, our numerics alone cannot rule out the possibility of a narrow critical MBL phase. To do this, and to confirm some of the suggestive numerical results presented in this section, in the next section we derive analytic expressions for the RG flow equations and for the form of the distribution of the lengths of thermal inclusions $\rho^T_\mathrm{dec} (L^T_\mathrm{dec})$.

\section{Analytics \label{sec:Analytics}}
In this section we further investigate our model of the MBL transition and adjacent phases, relying more on analytic rather than numerical methods. In particular, we present exact analytic flow equations for the RG flow parameters $f$ and $\tilde{\zeta}$ that confirm a KT-like RG flow for our model. We also derive the analytic form of the distribution of physical lengths of large rare thermal inclusions in the MBL phase and at the transition, showing that our model gives rise to stretched-exponential distributions in the MBL phase that become a power law only at the critical point, so there is no intermediate critical MBL phase.

\subsection{RG flow equations \label{subsec:RGflow2}}
Considering the RG rules of Section~\ref{sec:TheRG}, flow equations for the mean lengths $\braket{l}$, $\braket{L^I}$, and $\braket{L^T}$ can be obtained (see Appendix~\ref{app:Flowequationsmeans}). These, along with other results established later in this section, directly imply exact RG flow equations for $f$ and $\tilde{\zeta}$. They are
\begin{equation}
	\label{eqn:fTflow}
	\Lambda \frac{d f}{d\Lambda} = -\frac{ \Lambda \rho^T (\Lambda) - \braket{L^{I | d=\Lambda}} \mu^I (\Lambda)}{1+\Lambda \mu^I (\Lambda)} \frac{1-(1-f)\tilde{\zeta}}{\tilde{\zeta}}
\end{equation}
and
\begin{equation}
	\label{eqn:zetatildeflow}
	\Lambda \frac{d \tilde{\zeta}}{d\Lambda} = \frac{(\braket{L^{I | d=\Lambda}}+\Lambda) \mu^I (\Lambda)}{1+\Lambda \mu^I (\Lambda)} \Lambda \mu^I (\Lambda) (1-(1-f)\tilde{\zeta}),
\end{equation}
where $\rho^T(\Lambda)$ and $\mu^I(\Lambda)$ are the probability densities of $T$ and $I$ blocks with primary length equal to the cutoff $\Lambda$, and $\braket{L^{I | d=\Lambda}}$ is the average $L^I$ of $I$ blocks at the cutoff. 

This is different than the approach taken by the authors of~\cite{Zhang-Huse2016,Goremykina-Serbyn2018}, who directly consider the full integro-differential equations governing the evolution of the single-block probability distributions. Because of the extra complications in our RG, we choose not to follow that strategy, but the equations governing the flow of $\rho^T(L^T)$, $\rho^I(d,L^I)$, and the marginal distribution $\mu^I(d)$ are given in Appendix~\ref{app:Flowequationsdists}. We note, however, that the numerical simulations of Section~\ref{sec:Numerics} are essentially equivalent to evolving increasingly coarse representations of the distributions $\rho^T(L^T)$ and $\rho^I(d,L^I)$ according to these integro-differential equations.

In order to analyse the RG flow specified by Equations~(\ref{eqn:fTflow}) and~(\ref{eqn:zetatildeflow}), we first establish an understanding of how $\rho^T(\Lambda)$ and $\mu^I(\Lambda)$ flow. The flow equation for the entire marginal distribution $\mu^I(d)$ can be exactly reduced to an equation governing the flow of $\mu^I(\Lambda)$ using the ansatz
\begin{equation}
	\label{eqn:muIansatz}
	\mu^I (d) = \mu^I (\Lambda) \exp \left( - (d - \Lambda) \mu^I(\Lambda) \right),
\end{equation} 
where $\mu^I (\Lambda)$ obeys
\begin{equation}
	\label{eqn:muILambdaflow}
	\frac{d\mu^I (\Lambda)}{d\Lambda} = - \mu^I (\Lambda) \rho^T (\Lambda).
\end{equation}
This is an exact result; the flow is closed and stable within the subspace of exponentially distributed $d$. A useful flow equation for $\rho^T(\Lambda)$ is not similarly available, so we determine the behavior of $\rho^T(\Lambda)$ numerically. The data shown in Figure~\ref{fig:flow}\hyperref[fig:flow]{(c)} supports the scenario where in the MBL phase and at the critical point $\rho^T(\Lambda)=(1+\kappa) / \Lambda^{1-\epsilon}$, with $\kappa$ and $\epsilon$ continuously varying throughout the MBL phase and critical values consistent with $\kappa_\mathrm{c} \ge 0$ and $\epsilon_\mathrm{c} = 0$. Along with Equation~(\ref{eqn:muILambdaflow}), and in the approximation that $\kappa$ and $\epsilon$ don't flow or flow slowly, these observations imply that $\mu^I(\Lambda)\propto \exp(-\frac{1+\kappa}{\epsilon} \Lambda^\epsilon)$ in the MBL phase, and $\propto \Lambda^{-(1+\kappa_\mathrm{c})}$ along the critical flow. In the thermal phase $\mu^I(\Lambda)$ approaches a constant and $\rho^T(\Lambda)$ decays exponentially.

Returning to the analysis of Equations~(\ref{eqn:fTflow}) and~(\ref{eqn:zetatildeflow}), these equations and the flow of $\rho^T(\Lambda)$ and $\mu^I(\Lambda)$ detailed in this section imply that in the MBL phase the flow lines approach a fixed line at $f=0$ and $0<\tilde{\zeta}<1$, as suggested in~\cite{Dumitrescu-Vasseur2018}. Furthermore, the flow of $f$ and $\tilde{\zeta}$ are proportional to the factor $1-(1-f)\tilde{\zeta}$, implying a change of the stability of the fixed line at $f=0$, $\tilde{\zeta}=1$ where the MBL phase ends and a global avalanche instability occurs.  We therefore conclude that our RG does indeed exhibit a KT-type flow.

\subsection{Rare large thermal inclusions \label{subsec:Thermalinclusions2}}
The proposed critical MBL phase of~\cite{Goremykina-Serbyn2018,Dumitrescu-Vasseur2018} contains rare large thermal inclusions with fractal dimension zero, whose lengths are distributed as a power law.  Here we show that our model does not contain an intermediate MBL phase by showing that such an extended power law regime does not exist in our model, and arguing that the fractal dimension of large thermal inclusions decreases continuously in the MBL phase, reaching zero only at the single critical point in our RG, thus following ``scenario (i)'' of Ref.~\cite{Dumitrescu-Vasseur2018}.

\begin{figure}
		\includegraphics[width=0.8\linewidth]{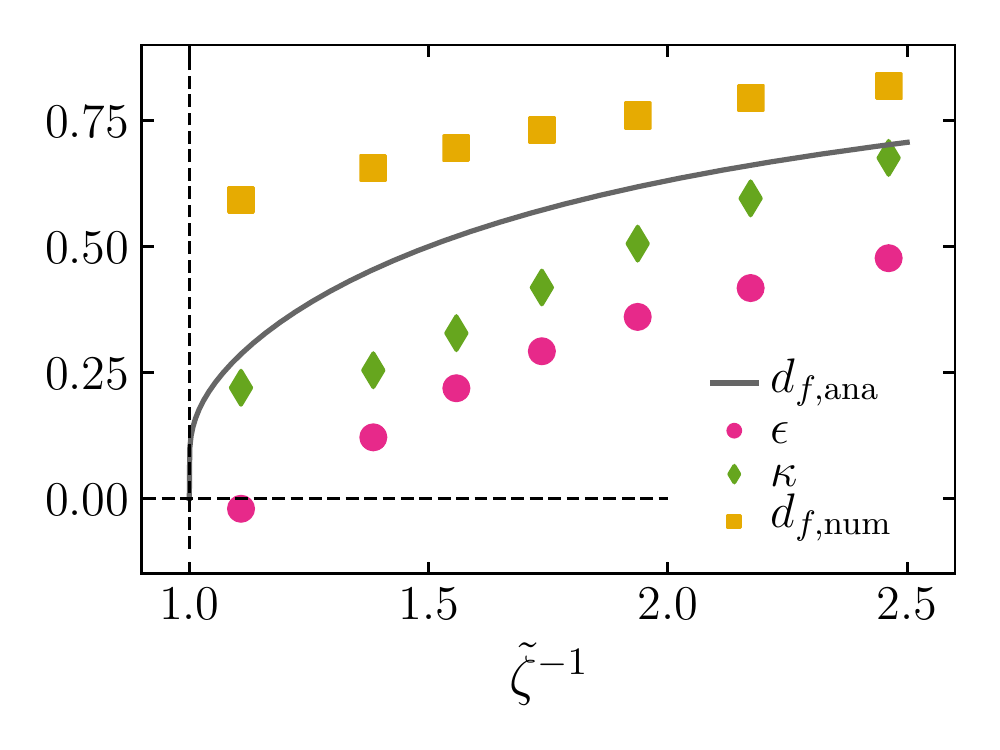}
	\caption{Numerical estimates of the parameters $\epsilon$ and $\kappa$, and the fractal dimension $d_f$ deep in the MBL phase and at criticality (from right to left). $\epsilon$ and $\kappa$ were estimated by fitting numerical values of $\rho^T(\Lambda)$ to the form $(1+\kappa)/\Lambda^{1-\epsilon}$. $d_{f,\mathrm{ana}}$ is the expression given in Equation~(\ref{eqn:fractaldim}). $d_{f,\mathrm{num}}$ was computed by fitting finite-size RG data to the form $\braket{L^T_\mathrm{frac}} \propto \braket{L^T}^{d_{f,\mathrm{num}}}$.  The points are plotted using the values of $\tilde\zeta$ obtained at the end of the finite-size simulations of the RG flows, which do not reach $\tilde\zeta=1$ for the critical systems. The discrepancy between the estimates $d_{f,\mathrm{num}}$ and $d_{f,\mathrm{ana}}$ can be accounted for by well-understood finite-size effects; this is discussed in the main text of Section~\ref{subsec:Thermalinclusions2}. \label{fig:exponents}}
\end{figure}

In Section \ref{subsec:Thermalinclusions1} we defined the cumulative distribution $\rho^T_\mathrm{dec}(L^T_\mathrm{dec})$ as the distribution of the physical lengths of $T$ blocks decimated by $ITI\mapsto I$ moves, and interpreted it to represent the distribution of thermal inclusions in large MBL regions. This cumulative distribution is given by $\rho^T_\mathrm{dec} (L^T_\mathrm{dec}) \propto \left. N_\Lambda \rho^T_{\Lambda}(\Lambda) \right|_{\Lambda=L^T_\mathrm{dec}}$ where $N_\Lambda$ is the number of blocks in the chain when the cutoff is $\Lambda$ and flows according to $\frac{dN}{d\Lambda} = -N(\mu^I(\Lambda) + \rho^T(\Lambda))$. Recalling from Section~\ref{subsec:RGflow2} that $\rho^T(\Lambda) = (1+\kappa) / \Lambda^{1-\epsilon}$ in the MBL phase ($\epsilon>0$, $\kappa>0$) and at criticality ($\epsilon_\mathrm{c}=0$, $\kappa_\mathrm{c} \ge 0$), this implies that as long as $\kappa >0$ the lengths of large thermal inclusions are distributed according to
\begin{equation}
	\label{eqn:rhoTdec}
	\rho^T_\mathrm{dec}(L^T_\mathrm{dec}) \propto \frac{\exp \left( -\frac{1+\kappa}{\epsilon} \left( (L^T_\mathrm{dec})^\epsilon - \Lambda_0^\epsilon \right) \right)}{(L^T_\mathrm{dec})^{1-\epsilon}},
\end{equation}
where $\Lambda_0$ is an integration constant. In the MBL phase this is a stretched exponential, and only at the critical point in the limit of $\epsilon \to 0$ does it take the form of a power law $\propto (L^T_\mathrm{dec})^{-\alpha}$, with $\alpha = 2+\kappa_\mathrm{c}$. In the case that $\kappa_\mathrm{c}$ does vanish at criticality, there is a subleading term not shown in Equation~(\ref{eqn:rhoTdec}) that allows for $\alpha=2$ only if $\lim_{\Lambda \to \infty} \Lambda \mu^I(\Lambda) = 0$ at the critical point. In our finite-size simulations we find $\kappa_\mathrm{c} \cong 0.2$, as demonstrated in Figures~\ref{fig:flow}\hyperref[fig:flow]{(c)} and~\ref{fig:exponents}, however this may not be the asymptotic value for infinite systems. We therefore conclude that our RG is consistent with $2 \le \alpha  \lesssim 2.2$. We have also confirmed that for $\epsilon$ and $\kappa$ determined by fitting $\rho^T(\Lambda)$ to the form $(1+\kappa)/\Lambda^{1-\epsilon}$ in numerical simulations, the form of Equation~(\ref{eqn:rhoTdec}) does indeed fit the numerical distribution of $L^T_\mathrm{dec}$. Most notably, our RG model does not contain an intermediate MBL phase where the physical length of thermal inclusions is distributed according to a power law. This property is only present at the critical point.

In order to obtain an estimate of the fractal dimension of large rare $T$ blocks in our RG, we follow the type of argument used in~\cite{Zhang-Huse2016,Goremykina-Serbyn2018} to compute the same quantity. Large $T$ blocks in the MBL phase are created via rare $TIT\mapsto T$ moves. $I$-blocks at the cutoff have physical length near $\frac{ \tilde{\zeta} }{ 1-\tilde{\zeta} } \Lambda$, and typical thermal blocks are of length close to $\Lambda$ in the MBL phase at large scales. Therefore, the typical size of large thermal inclusions resulting from rare $TIT\mapsto T$ moves is $\left( 2 +\frac{\tilde{\zeta}}{1-\tilde{\zeta}} \right) \Lambda$, while the total length of contributing thermal blocks is $2\Lambda$.  Accordingly, the largest rare $T$ blocks in the MBL phase are made from smaller $T$ blocks that form a fractal Cantor-like set having an asymptotic fractal dimension of
\begin{equation}
	\label{eqn:fractaldim}
	d_{f,\mathrm{ana}} = \frac{\log (2 )}{\log \left( 2 +\frac{\tilde{\zeta}}{1-\tilde{\zeta}} \right)},
\end{equation}
so $d_{f,\mathrm{ana}} \to 0$ continuously only as $\tilde{\zeta} \rightarrow 1$ as the critical point is approached along the MBL fixed line, and this approach of $d_{f,\mathrm{ana}}$ to zero is logarithmically slow in $1-\tilde\zeta$. 

The fractal structure of $T$ blocks in the MBL phase implies that, during finite-size numerical simulations of the RG, $\braket{L^T_\mathrm{frac}} \propto \braket{L^T}^{d_{f,\mathrm{num}}}$, where $L^T_\mathrm{frac}$ is the total length of the parts of a $T$ block that remained in $T$ blocks throughout the entire RG, and $d_{f,\mathrm{num}}$ is a numerical estimate of the fractal dimension of large $T$ blocks. We use this implication to numerically compute the fractal dimension $d_{f,\mathrm{num}}$ from finite-size simulations and the two estimates, $d_{f,\mathrm{ana}}$ and $d_{f,\mathrm{num}}$, are shown in Figure~\ref{fig:exponents} for systems throughout the MBL phase and at criticality. There is a strong finite-size effect relevant to $d_{f,\mathrm{num}}$ that can be seen in Figure~\ref{fig:exponents}:  In finite-size simulations the typical length of $T$ blocks in the MBL phase is actually larger than $\Lambda$, which causes the value of $d_{f,\mathrm{num}}$ to overestimate the asymptotic value of the fractal dimension.  This effect increases as one approaches the critical point. Overall, these estimates of $d_f$ give no indication of a possible intermediate MBL phase with $d_f=0$.

As a final note, it has been suggested that the stretching exponent $\epsilon$ of Equation~(\ref{eqn:rhoTdec}) for the $T$ block length distribution and the fractal dimension $d_f$ are equivalent quantities~\cite{Agarwal-Knap2017}. This is because in order to grow a large thermal region of length $L^T$ it seems that the number of independent rare events that must occur at smaller scales is $\propto (L^T)^{d_f}$.  However, the rarity of those events does change with scale, so precisely how to estimate $\epsilon$ by this route is not clear. In Figure~\ref{fig:exponents} we show numerical estimates of $\epsilon$ and $d_f$, as we have defined and computed them, and find they are quite different in the MBL phase. Comparing different system sizes does not suggest that this difference is due to finite-size effects.  Thus at this point it appears that these are independent exponents with rather different dependences on $\tilde\zeta$, although we do expect that they both go asymptotically to zero at the critical point.

\section{Conclusions \label{sec:Conclusions}}
By building on the RGs of~\cite{Zhang-Huse2016,Goremykina-Serbyn2018} we deviated from complete analytic solvability to include some of the physics of quantum avalanches into an approximate RG description of the MBL phase and MBL transition. This was done by giving each $I$ block a decay length $\zeta$ for l-bit-flip interactions within the block. Using numerical and analytic approaches we studied the RG flow of our model and concluded that our RG exhibits a KT-type flow. This suggests that the KT-type MBL transition, where critical systems are localized and at the precipice of a global avalanche instability at the largest scales, may be valid~\cite{Goremykina-Serbyn2018,Dumitrescu-Vasseur2018}. In our RG the MBL phase is parametrized by a global decay length $\tilde{\zeta}$, with fractal thermal inclusions whose lengths are distributed as a stretched exponential, and fractal dimension that varies within the phase. Only at the critical point is the distribution of the physical lengths of thermal inclusions a power-law with exponent $\alpha \cong 2$, and their fractal dimension approaches zero.  Therefore, our model of the MBL phase and phase transition in one dimension does not generate support for an intermediate critical MBL phase.

\begin{acknowledgments}
We would like to thank Wojciech De Roeck, Philipp Dumitrescu, Sarang Gopalakrishnan, Anya Goremykina, Vedika Khemani, Sid Parameswaran, Maksym Serbyn and Romain Vasseur for insightful discussions.  AM acknowledges the support of the Natural Sciences and Engineering Research Council of Canada (NSERC).  DAH was supported in part by the DARPA DRINQS program.
\end{acknowledgments}

\bibliography{newRG_PRB_submission}

\appendix

\section{Flow equations for mean lengths \label{app:Flowequationsmeans}}
The flow equations for $\braket{l}$, $\braket{L^I}$, and $\braket{L^T}$ are
\begin{eqnarray}
	\frac{d\braket{l}}{d\Lambda} &=& (\braket{l}-\Lambda-\braket{L^{I | d=\Lambda}})\mu^I (\Lambda) \nonumber \\ 
	& & + \braket{l}\rho^T (\Lambda) \label{eqn:meanlflow} , \\
	\frac{d\braket{L^I} }{d\Lambda} &=& (\braket{L^I} - \braket{L^{I | d=\Lambda}})\mu^{I}(\Lambda) \nonumber \\
	& & + (\braket{L^I}+\Lambda)\rho^T(\Lambda) \label{eqn:meanLIflow} , \\
	\frac{d\braket{L^T}}{d\Lambda} &=& (\braket{L^T} - \Lambda)\rho^T (\Lambda) \nonumber \\
	& & + (\braket{L^T}+\braket{L^{I | d=\Lambda}})\mu^{I} (\Lambda) \label{eqn:meanLTflow} ,
\end{eqnarray}
where $\braket{L^{I | d=\Lambda}}$ denotes the mean physical length of $I$ blocks at the cutoff. In this Appendix we derive Equation~(\ref{eqn:meanLTflow}). The derivation of Equations~(\ref{eqn:meanlflow}) and~(\ref{eqn:meanLIflow}) are similar.

$ITI\mapsto I$ and $TIT \mapsto T$ moves have three types of effects on the flow of $\braket{L^T}$ herein denoted types $a$,$b$, and $c$. The first (type $a$) arises with the decimation of $T$ blocks having $L^T=\Lambda$ in $ITI\mapsto I$ moves. This happens with weight $\rho^T(\Lambda)d\Lambda$ during a small increase of the RG scale from $\Lambda$ to $\Lambda+d\Lambda$. After only type $a$ effects are accounted for the mean $L^T$ is
\begin{eqnarray}
	\braket{L^T}_{a} = \frac{ \braket{L^T} -(\rho^T (\Lambda) d\Lambda) \Lambda }{1 - \rho^T (\Lambda) d\Lambda}.
\end{eqnarray}
Type $b$ effects come from the decimation of $T$ blocks with any value of $L^T$ in $TIT\mapsto T$ moves, which happens at a rate proportional to $2 \mu^I(\Lambda) d\Lambda$. Type $b$ effects do not alter the mean $L^T$, \textit{i.e.} $\braket{L^T}_{a,b} = \braket{L^T}_{a}$. Type $c$ effects come from the creation of new $T$ blocks, which happens at a rate proportional to $\mu^I(\Lambda) d\Lambda$. The average $L^T$ of a newly merged $T$ block when the cutoff is $\Lambda$ is $\braket{L^T_\mathrm{new}} = 2\braket{L^T} + \braket{L^{I | d =\Lambda}}$. After all three effects are accounted for, the mean $L^T$ is 
\begin{equation}
	\braket{L^T}_{a,b,c} = \frac{ (1-2\mu^I  (\Lambda) d\Lambda) \braket{L^T}_{a,b} + (\mu^I (\Lambda) d\Lambda) \braket{L^T_\mathrm{new}} }{1 - \mu^I(\Lambda) d\Lambda}.\nonumber
\end{equation}
This quantity is better-denoted $\braket{L^T}_{\Lambda+d\Lambda}$ now. Substituting in the given expressions for $\braket{L^T}_{a,b}$ and $\braket{L^T_\mathrm{new}}$ and taking the infinitesimal limit of $d\Lambda$ yields the flow equation for $\braket{L^T}$, Equation~(\ref{eqn:meanLTflow}).

\section{Flow equations for single-block distributions \label{app:Flowequationsdists}}
In this Appendix we provide the full integro-differential equations for the single-block distributions $\rho^T(L^T)$ and $\rho^I(d,L^I)$, which fully determine the RG flow. We also include the equation for $\mu^I(d)$, the marginal distribution.
\begin{widetext}
	\begin{eqnarray}
		\frac{\partial \rho^T}{\partial \Lambda} =&& (\rho^T (\Lambda)-\mu^I (\Lambda))\rho^T  + \frac{L^T}{\Lambda}\frac{\partial \rho^T }{\partial L^T} \nonumber \\
		&&+ \mu^{I} (\Lambda) \Theta(L^T-2\Lambda) \int_\Lambda^{L^T-\Lambda} \int_\Lambda^{L^T-L^T_2} \rho^T (L^T_1) \frac{\rho^I (\Lambda,L^T-L^T_1-L^T_2)}{\mu^{I} (\Lambda)} \rho^T (L^T_2) dL^T_1 dL^T_2 \label{eqn:muTflow},\\
		\frac{\partial \rho^I }{\partial \Lambda} =&& (\mu^{I} (\Lambda)-\rho^T (\Lambda))\rho^I + \frac{d}{\Lambda}\frac{\partial \rho^I}{\partial d} + \frac{L^I}{\Lambda}\frac{\partial \rho^I}{\partial L^I}\nonumber \\
		&&+ \rho^T(\Lambda) \Theta(d-\Lambda) \Theta(L^I-\Lambda) \int_\Lambda^{d} \int_0^{L^I-\Lambda} \rho^I (d-d_1+\Lambda,L^I-L^I_1-\Lambda) \rho^I (d_1,L^I_1) dL^I_1 dd_1 \label{eqn:rhoIflow},\\
		\frac{\partial \mu^I}{\partial \Lambda} =&& (\mu^{I} (\Lambda)-\rho^T (\Lambda))\mu^I  + \frac{d}{\Lambda}\frac{\partial \mu^I}{\partial d} + \rho^T (\Lambda) \Theta(d-\Lambda) \int_\Lambda^{d} \mu^I (d-d_1+\Lambda) \mu^I (d_1) dd_1 \label{eqn:muIflow}.
	\end{eqnarray}
\end{widetext}

\end{document}